\documentclass[conference]{IEEEtran}
\usepackage{amsmath}
\usepackage{color}
\usepackage{amsthm} 
\usepackage{amssymb}
\usepackage{float}
\usepackage{nccmath}
\usepackage{graphicx}
\usepackage{kotex}
\usepackage[linesnumbered,ruled]{algorithm2e}
\usepackage{setspace}
\usepackage{booktabs}
\usepackage{subcaption}
\usepackage{mwe}
\usepackage{cite}
\usepackage{amsmath,amssymb,amsfonts}
\usepackage{graphicx}
\usepackage{textcomp}
\usepackage{xcolor}

\usepackage{algorithmicx}
\usepackage{xspace}

\begin{document}
\title{3D Modeling and WebVR Implementation 
using Azure Kinect, Open3D, and Three.js}

\author{\IEEEauthorblockN{$^{\circ}$Won Joon Yun and $^{\circ,\ddag}$Joongheon Kim}
\IEEEauthorblockA{$^{\circ}$School of Electrical Engineering, Korea University, Seoul, Republic of Korea 
\\$^{\ddag}$Artificial Intelligence Engineering Research Center, College of Engineering, Korea University, Seoul, Republic of Korea
\\
E-mails: \texttt{ywjoon95@korea.ac.kr}, 
\texttt{joongheon@korea.ac.kr}
}
}
\maketitle

\begin{abstract}
This paper proposes a method of extracting an RGB-D image using \textit{Azure Kinect}, a depth camera, creating a fragment, \textit{i.e.}, 6D images (RGBXYZ), using \textit{Open3D}, creating it as a point cloud object, and implementing webVR using \textit{three.js}. Furthermore, it presents limitations and potentials for development.
\end{abstract}

\IEEEpeerreviewmaketitle

\section{Introduction}\label{sec:intro}

Recently,the mobile device with LiDAR sensor, \textit{e.g.} Ipad Pro4, has been released and Apple has announced that it will release Apple Glass. After the VR industry emerged, it has been receiving a lot of attention from the academia and the public about virtual environments such as VR, AR, and MR around the world~\cite{iotj16lv,etri}.

Bringing the real world into virtual reality is another area of VR. The way to implement it is to use a sensor (\textit{e.g.} Microsoft's Azure Kinect, Intel's RealSense, etc.) to get an RGB-D image. The information obtained from the sensor is made into a mesh or point cloud through registration process which convert RGB-D image to point cloud (RGBXYZ) and data pre-processing using the platform((\textit{e.g.} Unity, Unreal Engine or Open3D). After that, rendering is done through a data rendering platform (\textit{e.g.} Hologram, Light Field, or VR). 

This paper supposes a method that restores the data received from the vision sensor into a complete point cloud and render it, using Azure Kinect as a vision sensor, preprocessing data using Open3D, and rendering it in WebVR.

\section{Sensing and Data-Processing}
\begin{figure*}[t!]
\centering
\includegraphics[page=1, width=1\linewidth]{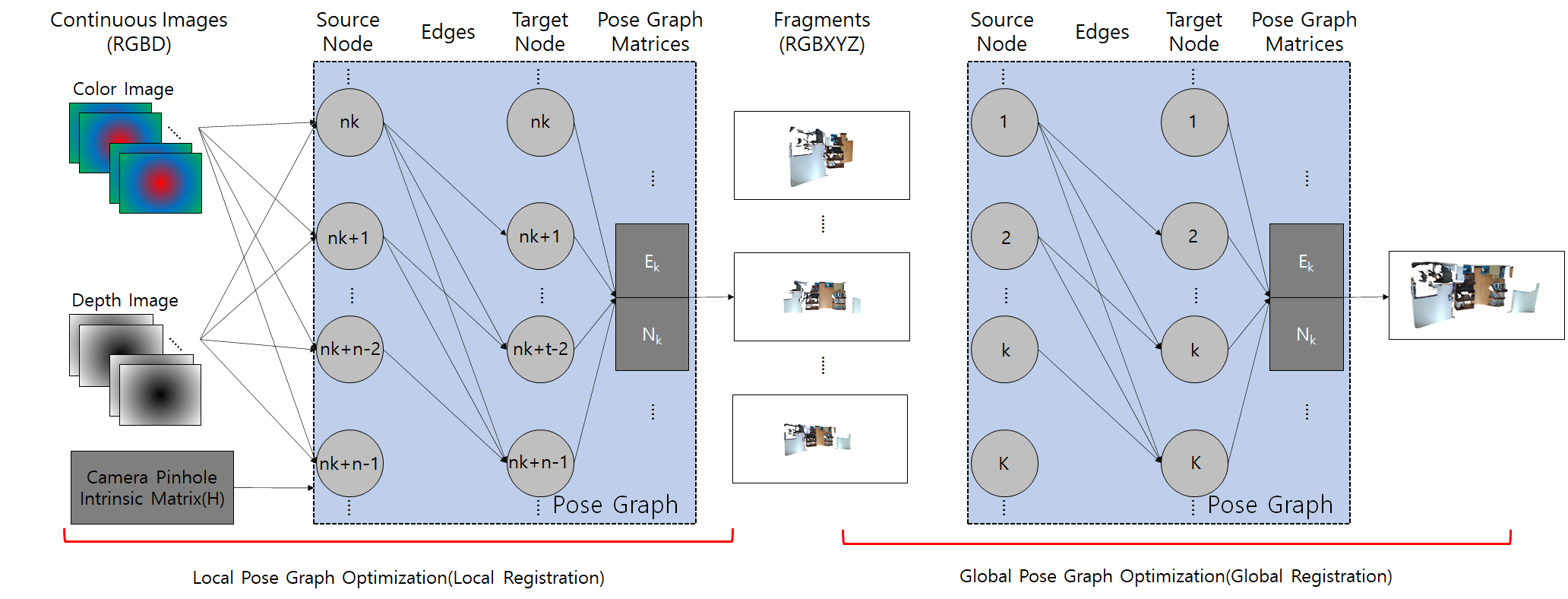}
\caption{Data processing overview.}
\label{dataprocess}
\end{figure*}

A video of a total of 240 frames was obtained by photographing a room, and this was extracted as a color image and a depth image pair.
Azure Kinect, an RGB-D sensor, was used for shooting, and data was obtained using Azure Kinect SDK.
Fig.~\ref{dataprocess} is a comprehensive representation of data processing.
From the extracted $240$ color images, depth images, and pinhole camera intrinsic matrix, $240$ point clouds are created and mapped to $240$ nodes.
Divide all the consecutive nodes by $N$ and group them. The nodes can be expressed as follows:
\begin{algorithm}[t!]
 \caption{ICP-Registration}
 \begin{algorithmic}[1]
    \State {\textbf{Input:} The set of RGB-D nodes $\mathcal{A} = \left\{ a_1, \dots, a_m , \dots, a_M \right\}$}, $N$ : the number indicating how many node pairs to divide.
    \State {\textbf{Output:} The set of point clouds $\mathcal{P} =  \left\{ p_1, \dots, p_k, \dots, p_K \right\}$}
    \Statex \textit{s.t.} $M = N\dot K$ \textit{where} $M,K$ and $N \in \mathcal{N}^1$
    \State \textbf{Initialize:} {$\mathcal{P} \leftarrow \left\{\right\}$, $\mathcal{T}=\left\{\right\}$}
    \State \textbf{Definition:} $s$: source node index, $t$: target node index \textit{s.t.} $ 1 \leq s < t < s + N \leq M$
    \Statex {}
    \State \textbf{for} $s$ in range $M$:
    \State \hspace{2em} $a_s$ : source node
    \State \hspace{2em} \textbf{for} $t$ in range $[s, \min(s+ N, M))$:
    \State \hspace{4em} $a_t$ : target node
    \State \hspace{4em} $\hat{a}_{s,t} \triangleq T_{s,t}\cdot a_t$
    \State \hspace{4em} $T^*_{s,t}\leftarrow \underset{T_{s,t}}{\arg\min}(|a_s-\hat{a}_{s,t}|)$
    \State \hspace{4em} $\mathcal{T} \leftarrow \mathcal{T}\cup T^*_{s,t}$
    \State \hspace{2em} \textbf{end}
    \State \textbf{end}
    \Statex {}
    \State \textbf{with} Jacobian RGBD odometry method(Input = $\mathcal{A}, \mathcal{T}$)
    \State \hspace{2em} \textbf{optimize} pose graph \textbf{then}
    \State \hspace{2em} \textbf{get} $\mathcal{P}$
    \State \textbf{end}
    \end{algorithmic}  
    \label{alg:icp}
\end{algorithm}

\begin{equation}
\nonumber
    \mathcal{A}=\left\{a_1  ,a_2,…,a_M \right\}
\end{equation} 

And for the above node set, suppose that source and target nodes exist. 
$a_s$ and $a_t$ represent a source node and a target node, respectively. $s$ and $t$ represent the indexes of the source node and the target node, respectively, which are following $1 \leq s < t < s + N \leq M$. 
For the source node $a_s$ and the target node $a_t$, there is an edge $T_{s,t}$ that satisfies as follow:
\begin{equation}
\nonumber
\hat{a}_{s,t} \triangleq T_{s,t} \cdot a_t
\end{equation}
And $\hat{a}_{s,t}$ can be obtained through the following equation:
\begin{equation}
\nonumber
T^*_{s,t} \leftarrow \underset{T_{s,t}}{\arg\min}(|a_s-\hat{a}_{s,t}|)
\end{equation}
\begin{equation}
\nonumber
\hat{a}_{s,t} \leftarrow T^*_{s,t} \cdot a_{t}
\end{equation}
where $T^*_{s,t}$ makes $|a_s- T_{s,t}\cdot a_t|$ minimum value.

The set of edges for all $a_s$ and $a_t$ can be expressed as follows:
\begin{equation}
\nonumber
    \mathcal{T}=\left\{T_{s,t} | 1 \leq s < t < s + N \leq M, s,t\in N^1 \right\}
\end{equation}

\begin{figure*}[h!]
\centering
\includegraphics[page=1, width=1\linewidth]{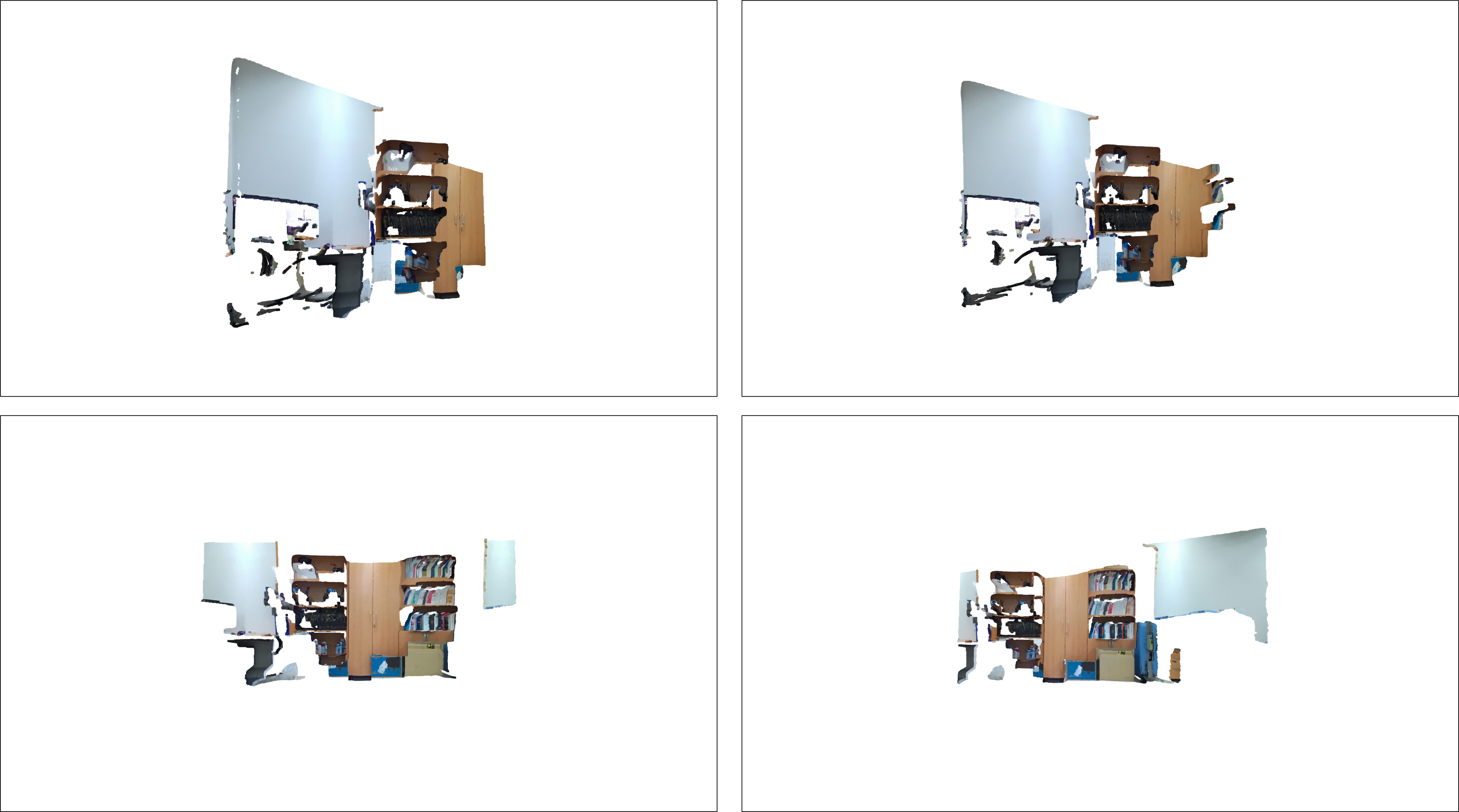}
\caption{Point cloud when local registration is completed.}
\label{fig:local}
\end{figure*}

Since the received RGB-D data is large, down-sampling is performed for each data in order to reduce the amount of computation to obtain the pose graph.
Then, proceed with pose graph optimization using \textit{Jacobian RGB-D odometry} method ~\cite{stein2011}.
When the optimal pose graph is obtained, a linear transformation operation is taken for each node and edge, and the position and phase are matched for $N$ consecutive point clouds. 
As a result, $K$ point clouds can be obtained where $K = \lceil240/N\rceil$. The set of point clouds is denoted as $\mathcal{P} =  \left\{ p_1, \dots, p_k, \dots, p_K \right\}$. And it can be shown in Fig ~\ref{fig:local}.
This process is called ICP-Registration.

\begin{figure*}[t!]
\centering
 \includegraphics[page=1, width=1.5\columnwidth]{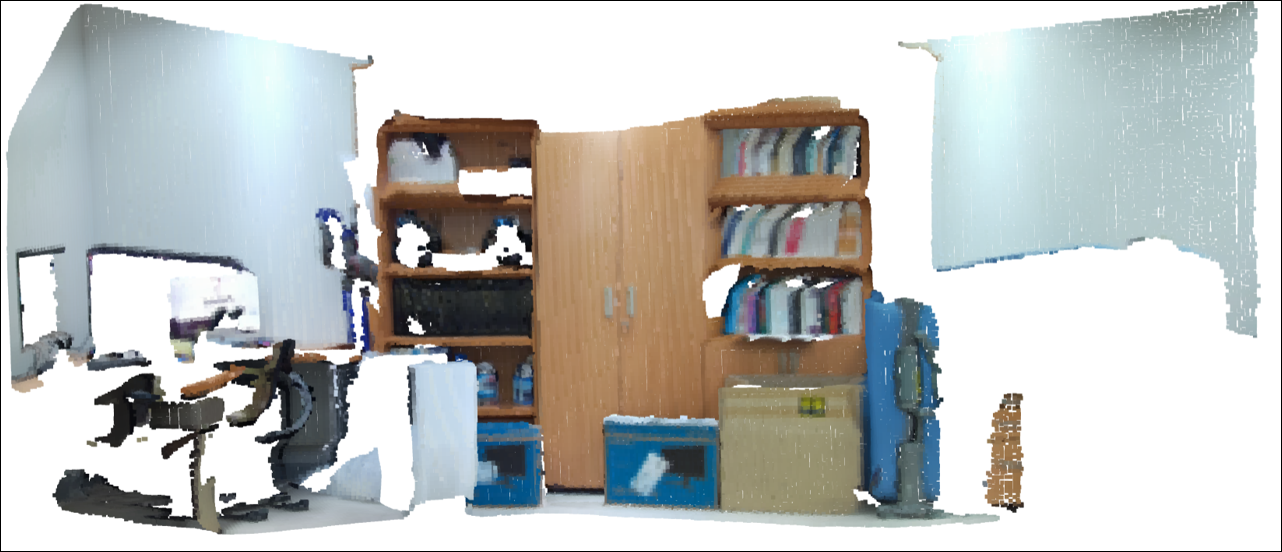} \\
\caption{The result of global registration.}
\label{fig:global}
\end{figure*}

\begin{figure*}[h!]
\centering
 \includegraphics[page=1, width=1.5\columnwidth]{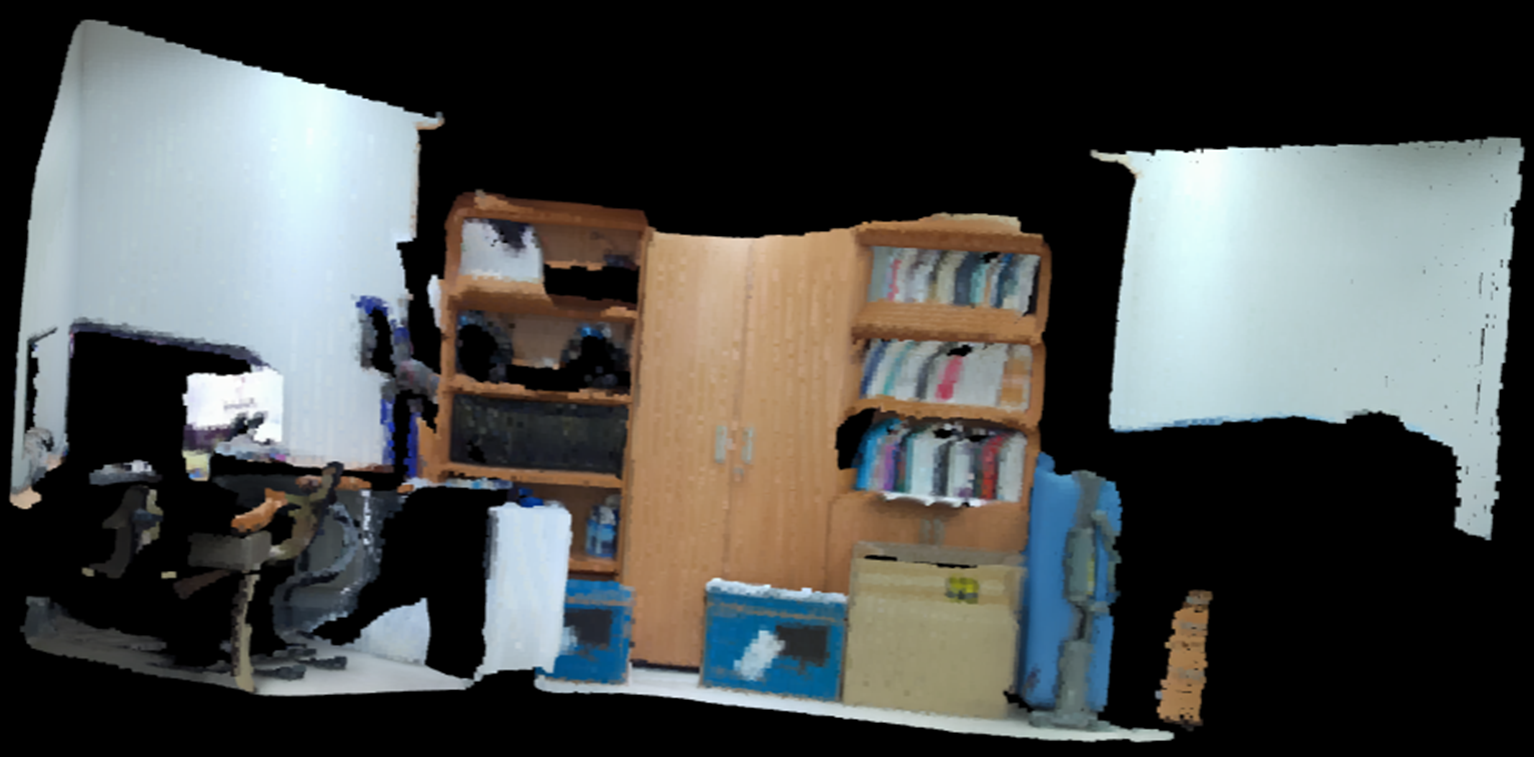} \\
\caption{The result of rendering the point cloud using \textit{Three.js}.}
\label{fig:vr}
\end{figure*}

After that, ICP-registration is performed for the $K$ point clouds $\mathcal{P}$ that have been locally registered in the similar way as above to match the location and image of the entire frame. 
For all natural numbers $k$ where $0<k<K$, define a possible set of $S_{k,k+1}\triangleq(p^s_k, p^t_{k+1})$ as follows.
\begin{equation}\nonumber
    \mathcal{S}\triangleq\left\{S_{1,2}, \cdots, S_{k,k+1}, \cdots, S_{K-1,K}\right\}
\end{equation}
For all $S_{k,k+1}$, calculate the transformation matrix $V_{k,k+1}$ which minimizes the distances between two point clouds $p^s_k$ and $p^t_{k+1}$. The set of transformation matrices is expressed as follows.
\begin{equation}
\nonumber
    \mathcal{V}\triangleq\left\{V_{1,2}, \cdots, V_{k,k+1}, \cdots, V_{K-1,K}\right\}
\end{equation}
A series of processes is called as global registration that obtains only one point cloud by performing pose graph optimization using the transformation matrix set $\mathcal{V}$ and the point cloud set $\mathcal{P}$ ~\cite{Zhou2014}~\cite{Zhou2016}.
And it can be checked through Fig ~\ref{fig:global}.
As a result, it is possible to obtain a 3D color map (\textit{i.e}, one completed point cloud) for the captured image.
The \textit{Jacobian RGB-D odometry} method was used for ICP-registration, and the process of ICP-registration is shown in Algorithm~\ref{alg:icp}. 

The tools/libraries used in the experiment are the python 3.6.5, OpenCV 4.2.0, numpy 1.19 and \textit{Open3D} 1.0.0 ~\cite{6631104}~\cite{Zhou2018}.

\section{VR Rendering}
\textit{Three.js} is a library that renders 3D images on a website ~\cite{noauthororeditor2015threejs}.
In addition, it supports image rendering for VR devices, through which WebVR can be implemented.
The point cloud can be loaded by through \textit{PCDLoader}, a built-in function of Three.js that can load pcd file. Then if \textit{VRButton}, a module that can render pcd file into VR device, is used, it is possible to show point cloud via VR.
Fig.~\ref{fig:vr} represents the result of rendering using \textit{Three.js}.

\section{Conclusions And Future Work}
We showed the process of receiving RGB-D values from Azure Kinect, obtaining a complete point cloud using \textit{Open3D}, and implementing WebVR through \textit{Three.js}.
As shown in Fig.~\ref{fig:global} and ~\ref{fig:vr}, the limitation of this paper is that we could not solve the empty space of point cloud.
If solving for empty spaces, and if a point cloud with high resolution can be obtained, it presents the possibility of development that can be applied to hologram and light field technologies.

\section*{Acknowledgment}
This research was supported by IITP grant funded by the Korea government (MSIP) (No. 2017-0-00068, A Development of Driving Decision Engine for Autonomous Driving using Driving Experience Information).
J. Kim is the corresponding author of this paper.
 
\bibliographystyle{IEEEtran}

\end{document}